\documentclass[twocolumn,twoside]{IEEEtran}
\usepackage{graphicx}

\begin{document}
\title{IP Over ICN Goes Live}

\author{
\IEEEauthorblockN{George Xylomenos\IEEEauthorrefmark{1}, 
Yannis Thomas\IEEEauthorrefmark{1}
Xenofon Vasilakos\IEEEauthorrefmark{1},
Michael Georgiades\IEEEauthorrefmark{2},
Alexander Phinikarides\IEEEauthorrefmark{2},\\
Ioannis Doumanis\IEEEauthorrefmark{3},
Stuart Porter\IEEEauthorrefmark{3},
Dirk Trossen\IEEEauthorrefmark{4},
Sebastian Robitzsch\IEEEauthorrefmark{4},
Martin J. Reed\IEEEauthorrefmark{5},
Mays Al-Naday\IEEEauthorrefmark{5},\\
George Petropoulos\IEEEauthorrefmark{6},
Konstantinos Katsaros\IEEEauthorrefmark{6},
Maria-Evgenia Xezonaki\IEEEauthorrefmark{6} and
Janne Riihij\"{a}rvi\IEEEauthorrefmark{7}
}\\
\IEEEauthorblockA{\IEEEauthorrefmark{1}Athens University of Economics and Business, Greece}
\IEEEauthorblockA{\IEEEauthorrefmark{2}PrimeTel PLC, Cyprus}
\IEEEauthorblockA{\IEEEauthorrefmark{3}CTVC Ltd, UK}\\
\IEEEauthorblockA{\IEEEauthorrefmark{4}InterDigital Europe, UK}
\IEEEauthorblockA{\IEEEauthorrefmark{5}University of Essex, UK}
\IEEEauthorblockA{\IEEEauthorrefmark{6}Intracom SA Telecom Solutions, Greece}\\
\IEEEauthorblockA{\IEEEauthorrefmark{7}RWTH Aachen University, Germany}
}

\maketitle

\begin{abstract}
Information-centric networking (ICN) has long been advocating for radical changes to the IP-based Internet. However, the upgrade challenges that this entails have hindered ICN adoption. To break this loop, the POINT project proposed a hybrid, IP-over-ICN, architecture: IP networks are preserved at the edge, connected to each other over an ICN core. This exploits the key benefits of ICN, enabling individual network operators to improve the performance of their IP-based services, without changing the rest of the Internet. We provide an overview of POINT and outline how it improves upon IP in terms of performance and resilience. Our focus is on the successful trial of the POINT prototype in a production network, where real users operated actual IP-based applications. 
\end{abstract}


\section{Introduction}

Information-Centric Networking~(ICN)~\cite{icn_survey} has so far taken a clean-slate approach to network architecture, focusing on information exchange rather than on the endpoint-based communication of the current Internet. ICN has created an active research community~\cite{DONA, CCN, PURSUIT}, but has made little progress in replacing IP, as this would require not only standardization, but also agreement among too many stakeholders: operators, vendors, software developers, device makers and policymakers. 

Understanding the complexity of such a task, the partners of the H2020 POINT project used their experience with ICN projects, including PSIRP, PURSUIT and COMET, to bring a more pragmatic approach to the table. POINT proposed harnessing the innovation potential of IP-based applications and solutions, while benefitting from specific ICN solutions to improve performance and resilience~\cite{eucnc}. The focus of POINT is on an individual network operator who desires to offer better IP-based services by introducing ICN in its core network, without requiring changes to the rest of the Internet.

This paper first provides an overview of the architecture and goals of POINT. Then, it describes the successful closed trial of the POINT technologies over the network of project partner PrimeTel, with real users accessing IP-based applications, including HLS-based video streaming and IPTV. We show how POINT was able to improve upon IP in terms of performance and resilience, based on data gathered during the trial. Finally, we outline the open trial of the project which will take place in user homes served by PrimeTel's network.

\section{The POINT architecture}

The POINT architecture aims to replace the network of an individual operator, so as to improve the IP-based services offered to its customers. POINT is a drop-in replacement for the existing network: it does not require changes to existing User Equipment~(UE), or to the IP routers of other operators. This is achieved by combining an ICN core network with a set of Network Attachment Points~(NAPs), which reside at the periphery of POINT, serving as gateways between IP and ICN.

The baseline POINT architecture was derived from the PURSUIT architecture~\cite{PURSUIT}, where the UEs publish and subscribe to named information items. This publish/subscribe architecture is facilitated by three core functions: a Rendezvous~(RV) function that matches publishers and subscribers; a Topology Management~(TM) function that calculates paths between the nodes and encodes them into Forwarding Identifiers~(FIDs); and, a Forwarding Node~(FN) function that allows messages to be forwarded in the network based on the FIDs. The FIDs represent the set of links that a packet must traverse, whether a unicast path or a multicast tree. These FIDs are included in packet headers, allowing FNs to forward packets with a few bitwise operations, without requiring routing tables or any routing state. Consequently, the PURSUIT ICN architecture supports stateless multicast switching and native anycast.

\begin{figure}
\centering
\includegraphics[width=3.25in]{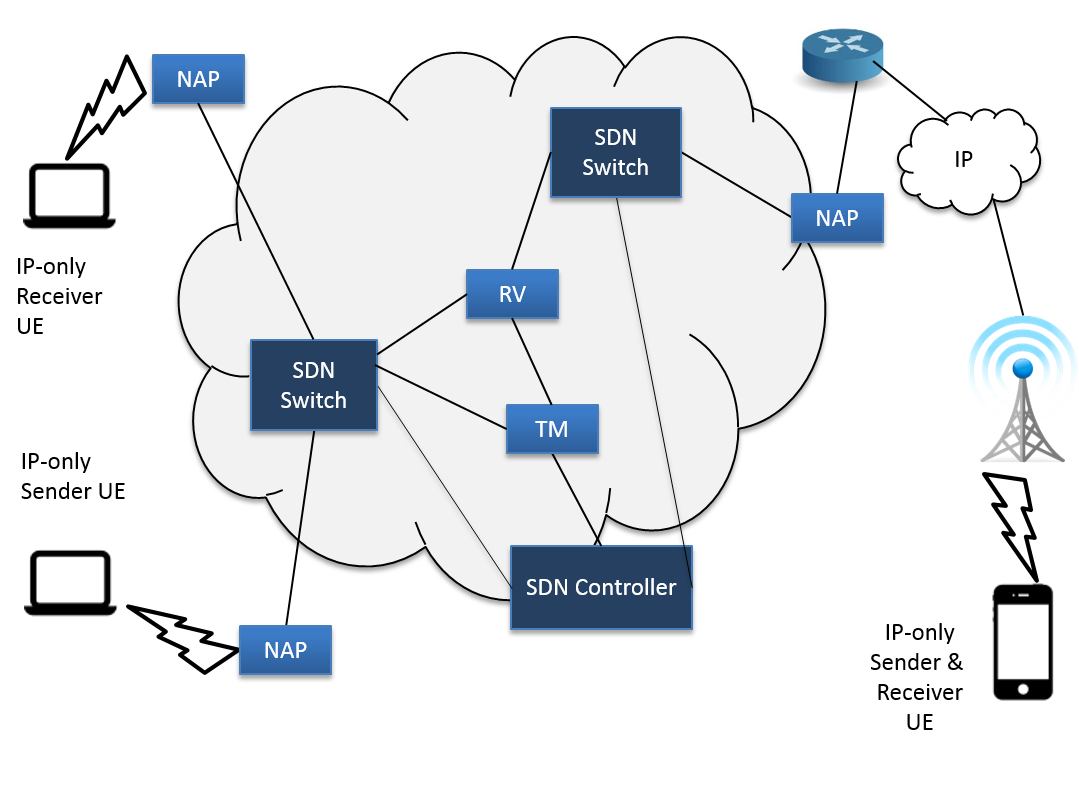}
\caption{The POINT architecture.}\label{architecture}
\end{figure}

Figure~\ref{architecture} outlines the main components of the POINT architecture, showing the physical connections between the various entities. The RV and TM functionalities are the main control functions of the ICN cloud -- in practice they may be co-located to form a single Path Computation Entity~(PCE) function. Standard Software Defined Networking~(SDN) switches are used for the FN functionality, a new feature introduced in POINT to simplify deployment in production networks. The SDN switches are unaware of POINT: they are controlled by an SDN controller~\cite{SDN} which communicates with the TM function in a bidirectional fashion. The TM instructs the SDN controller how to configure the SDN switches so as to translate the FIDs included in packets to forwarding actions on their attached links, while the SDN controller informs the TM of any changes in the topology and operation of the network.

Another new feature introduced by POINT is the fast formation of multicast trees, which is particularly important for video applications. In PURSUIT, adding or dropping a multicast receiver required communication with the TM which formed a new multicast tree. POINT takes advantage of the fact that the forwarding scheme allows forming a multicast tree by using a simple bitwise operation across its constituent unicast paths; thus, multicast senders cache unicast paths to receivers and dynamically combine them into multicast trees. As cached paths must be invalidated when the network topology changes, POINT developed a network monitoring scheme that allows network changes to be quickly communicated to all interested parties, thus allowing paths to be recalculated when needed.
 
To preserve the IP interface towards UEs and other operators, POINT uses a gateway approach. The NAPs, which are the access gateways of customers to the network, or the border routers/gateways to peering networks, handle all the protocols offered at the IP interface, either directly at the IP layer, or, if possible, at higher protocol layers, such as HTTP. As a result, the POINT network looks like a standard IP network to UEs and peering networks. 

\section{The goals of POINT}
 
In order to verify whether POINT does indeed offer a better service than an IP core, thus offering an incentive for adoption to network operators, we must first define what \emph{better} means~\cite{eucnc}. While the project identified and measured a large set of Key Performance Indicators~(KPIs), we will focus here on the performance and resiliency benefits for video delivery services, which were showcased in the closed POINT trial.

Current IP networks offer video delivery services using two different solutions. On the one hand, IPTV services offered by operators over their own access networks exploit IP multicasting to offer synchronized content viewing, relying on (statefull) snooping IGMP switches to support the economical distribution of many channels with fast channel switching. On the other hand, over-the-top services (e.g., YouTube) use adaptive HTTP streaming (e.g., HTTP Live Streaming, HLS) to cross diverse networks. Operators also use HLS to extend their IPTV services to customers currently attached to other networks, for example, 3G/4G cellular networks. Both approaches have shortcomings: the former requires a network with multiple non-standard extensions (e.g., IGMP snooping), while the latter requires individual users to separately stream the same content, even for live events. 

In the \emph{performance improvement} front, POINT can transparently combine independent HLS sessions of quasi-synchronized users, merging their individual requests as they enter the POINT network, and serving them with a single response delivered via multicast, which is turned back into individual unicasts when exiting the POINT network. This \emph{coincidental multicast} approach, where multicast groups are spontaneously formed when many requests happen to match, provides a dramatic reduction in network load, making solutions such as HLS competitive with standard IPTV solutions, as the overhead and redundancy of individual video streams vanishes. In principle, this could allow an operator to use HLS for both over-the-top services and native IPTV services.

In the \emph{service resilience} front, the reliance of POINT on centralized path management allows network operators to quickly adapt to changing network conditions. Paths can be recalculated at a single spot (the TM) and they only need to be communicated to the entry points of the POINT network to modify routing. In contrast, in IP networks routing is distributed, so changes must be slowly propagated everywhere before routing converges. For example, a scenario tested during the closed POINT trial switched HLS users from a failed HLS server (or path) to another server, without disturbing the service; the equivalent task in IP requires DNS-level changes which are slow and thus visibly disrupt HLS services.

Since IPTV services are so prevalent in operators' networks, POINT also developed a solution for supporting IPTV services based on IP multicasting and IGMP. The POINT solution has similar bandwidth requirements and latency performance with an IP-based solution, but it only relies on stateless SDN switches rather than statefull IGMP switches. The combination of stateless switches, centralized path management and source routing in POINT allows near instantaneous rerouting of multicast trees when paths fail. A scenario tested during the closed POINT trial switches all traffic from a failed trunk link to an alternative live trunk, without disturbing the IPTV service; the equivalent task in IP requires the switches to first rebuild their spanning tree and then rebuild their IGMP snooping state, visibly disrupting the IPTV service.

\section{The closed POINT trial}

\subsection{Overview}

From the outset of the POINT project, the goal was to test our prototype platform in a trial, conducted in the operational network of PrimeTel in Cyprus, so as to test the POINT prototype, refine it for operational deployment and exhibit its potential in a real ISP environment. A closed trial was concluded in late 2017, using the operational network of the company and its actual video services, with participants viewing content at PrimeTel's headquarters under our supervision.

In the first part of the trial, users were asked to watch a set of movie trailers distributed via HLS, first over the regular IP infrastructure of PrimeTel, and then over a POINT-enabled version of the network. We applied a series of exceptional circumstances to the network by simulating extreme network congestion towards the HLS server. Over IP the quality and resolution of the content was reduced, with visible artifacts, while over POINT the content appeared unaffected, as the network simply shifted the streams to another HLS server.

In the second part of the trial, users were asked to watch live television channels served over IPTV for a few minutes -- again, the content first travelled over a traditional IP network and then over a POINT-enabled network. This time, we applied exceptional circumstances by simulating a link being broken and then repaired in the network. Over IP, when the link failed, content transmission was interrupted, taking several minutes to switch to a backup link and the same occurred when the original link was restored. In contrast, over the POINT-enabled network, the content was immediately rerouted in both cases and the users did not experience any service failures.

In addition to traditional techniques, such as interviews and questionnaires, to gather user responses, we also used EEG (electroencephalogram) headsets to read user brainwave patterns. This enabled us to measure how levels of frustration increase subconsciously when users are faced with the kinds of exceptional conditions they experienced during the trials.

\begin{figure}
\centering
\includegraphics[width=3.5in]{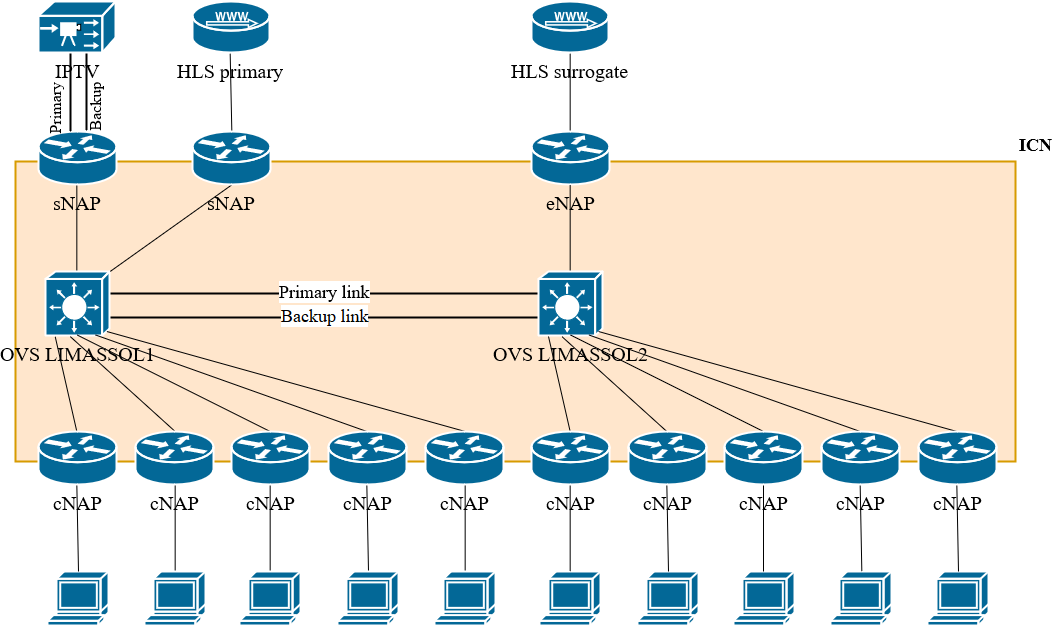}
\caption{Logical trial topology.}\label{topology}
\end{figure}

\subsection{Trial deployment}

Figure~\ref{topology} shows the logical topology implemented for the closed trial. The shaded area in the figure represents the POINT network, which is connected via a set of NAPs to regular IP clients and servers. On the top of the figure are three servers, one offering PrimeTel's IPTV service over UDP/IP mutlicast, and two (primary and surrogate) offering PrimeTel's TV Anywhere service over HLS, connected to server side NAPs~(sNAPs). On the bottom we have a number of clients, each connected to a client side NAP~(cNAP) via an ADSL network; the clients are either PrimeTel Set Top Boxes~(STBs) for IPTV or laptops for HLS. The sNAPs and cNAPs are connected to two SDN switches interconnected via two trunk links (primary and backup). We used Open vSwitch for the SDN switches, controlled by an OpenDaylight controller (not shown). The POINT software prototype used in the NAPs ran on regular Debian Linux 8.

This topology is a simplified version of actual ISP topologies that span multiple cities: in each city there are one or more distribution networks, with a number of customers downstream and, possibly, some servers upstream. Normally, customers are served by local servers, if available, but when such servers fail, they are served via the interconnection trunks from servers on other networks. It should be noted that the video servers are the actual production servers used by PrimeTel and the access network is PrimeTel's actual production ADSL network. The core network used for the closed trial is PrimeTel's R\&D network. For comparison purposes, the POINT network is running side-by-side with a regular IP network with the exact same topology, using VLANs to allow both networks to operate all the way from the servers to the client devices.

\subsection{Video Services: HLS}

PrimeTel uses HLS to deliver live video on top of the Internet when IP multicast cannot be used, which is the case when users are not connected to PrimeTel's distribution network. This is a common choice for operators serving customers connected to different networks, as well as for over-the-top providers, since HLS only requires HTTP, which is available universally. On the server side, a plain HTTP server is combined with a streamer/encoder to segment the video and encode the chunks into a range of qualities. On the client side, a (typically Javascript-based) application decides on the quality of the video to fetch, adapting to network conditions. Since each user is served over an individual HTTP session, a farm of HLS servers is needed to serve large numbers of customers. When live video is transmitted, segmentation and encoding must take place in real time, thus the HLS server needs to rewrite periodically the playlist files which define the available video qualities and the list of available video chunks.

The first goal of POINT for HLS services is to avoid unicast traffic by exploiting coincidental multicast for HTTP. Since HLS clients choose the quality of the next video chunk depending on their individually perceived network conditions, there is no way to predict their behaviour. Instead, multicast groups are formed dynamically as requests arrive, exploiting POINT's capability for very fast multicast group formation. The HTTP handler adds a slight delay when responding to HLS requests, to increase the chance of finding more similar requests to serve as part of a multicast transmission.

In addition to saving bandwidth, POINT supports quick switching of HLS clients to their closest server, relying on its centralized topology management and the need to only update routing information at the entry points of the network. To test this, we have introduced two identical HLS servers in the trial network, connected to two different NAPs, a primary and a surrogate server. A surrogate agent manually introduces the surrogate server on demand; this could be due to failure in the primary server or the path to it, or due to the need to balance the load on the network or on the video servers. The clients do not notice anything when such events occur. In contrast, in an IP network switching to a different server either requires hiding all servers behind a load balancer (which is not easy if they are not co-located) or assigning multiple IP addresses to the same DNS name, and letting clients switch to an alternative IP address when their current one fails, a very slow process.

\subsection{Video Services: IPTV}

The IPTV service offered by PrimeTel is delivered via IP multicast, using RTP and UDP to transport the video stream. While IP multicast is straightforward to implement over broadcast networks, it is far harder to implement on an ADSL access network, even though its tree-like structure is ideally suited to multicast. To implement IP multicast, the access network normally relies on IGMP snooping switches which analyze IGMP messages to locate which port a stream is originating from and which ports it should be forwarded to, rather than broadcasting each IP multicast stream everywhere. This means that IGMP snooping switches rely on state to economically forward multicast streams over the access network.

The POINT network also supports native multicast, but with stateless SDN switching~\cite{SDN}. Multicast trees are formed at the sNAPs attached to the stream sources, using the centralized TM to create paths and the fast multicast formation scheme to quickly add and drop recipients. Since source routing is used, there is no state to maintain and communicate to the SDN switches. While this approach is no faster or more economical than what an IPTV service already offers, it is very beneficial when failures occur. In practical networks, there are multiple paths between some switches for redundancy; we have indeed introduced this aspect in the closed trial network, interconnecting the two SDN switches with two trunks.

In a regular IPTV network, a spanning tree is created over the topology and some links are disabled for traffic forwarding purposes to avoid loops. When a link fails, the switches attached to it trigger a recalculation of the spanning tree. Due to the topological change however, the state of the IGMP snooping switches is now partially invalid, as it reflects a tree that does not exist. As a result, IGMP state needs to be re-established by snooping inside new IGMP messages, a time consuming process. In contrast, in the POINT network, there is no need for spanning trees: both links can be used all the time, with the centralized TM allocating different routes to each one. If one link goes down, only the NAP attached to the IPTV server must modify its routes to switch to the other.

\subsection{Monitoring the network}

\begin{figure}
\centering
\includegraphics[width=3.5in]{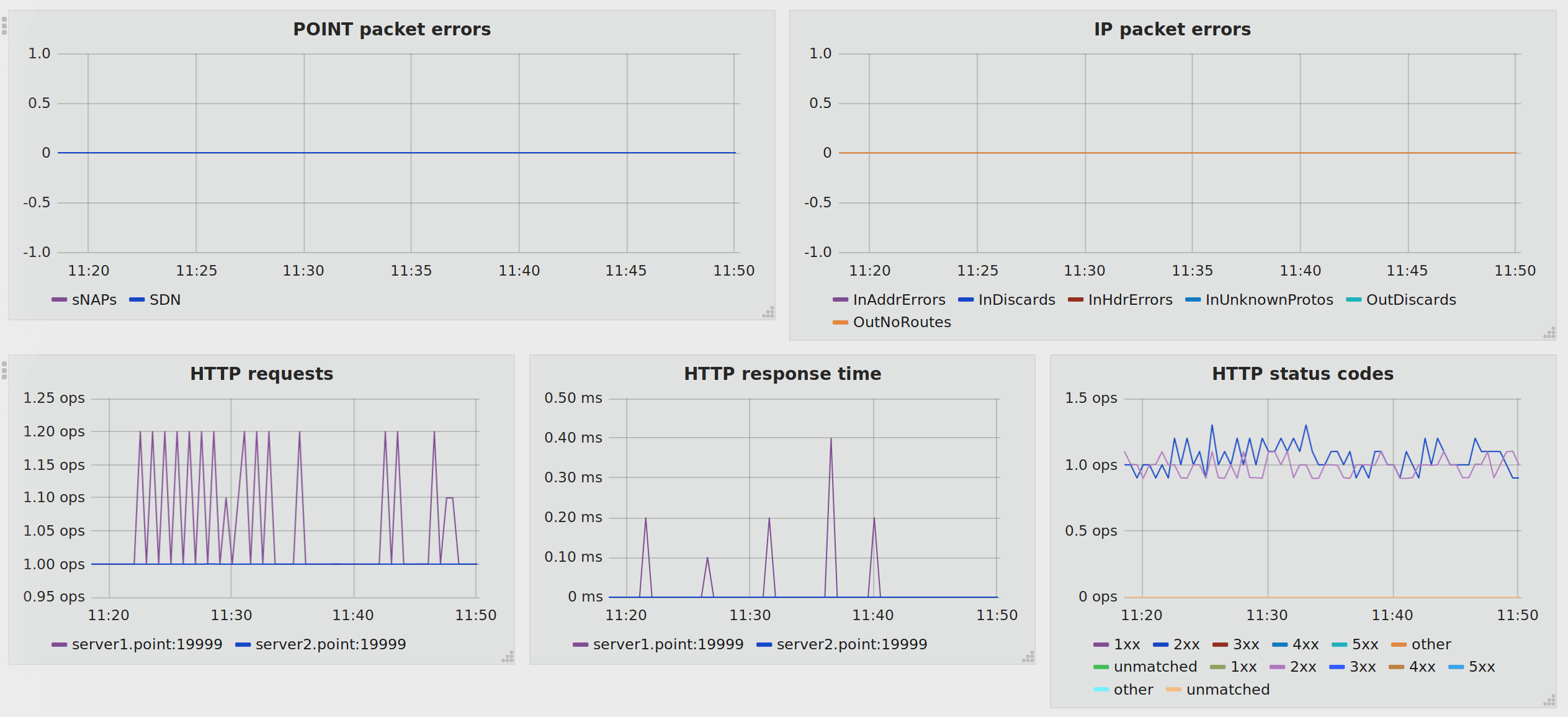}
\caption{A network visualization panel from the closed trial.}\label{visualization}
\end{figure}

Since POINT is aimed to network operators, it must address operator needs in terms of monitoring and visualization tools. We have therefore created a monitoring library that reports statistics to a local monitoring agent in each POINT element. A monitoring server queries these agents to gather, log and visualize monitoring data. A rich set of monitoring points has been defined, which can easily be extended (and was indeed greatly extended for the needs of the trial). For example, the NAPs monitor IPTV related statistics such as received and transmitted bytes and channel acquisition times. We have used existing monitoring solutions to gather additional monitoring data from non-POINT devices, including STBs, servers and IP-based networking devices, such as operating system, per process/network interface and HTTP-level metrics. As an example, Figure~\ref{visualization} shows a visualization panel with data on network and application level performance, for example, packet errors, HTTP requests, response times and status codes.

\subsection{Trial execution}

Between November 20th and December 1st 2017, we conducted the closed trial in PrimeTel's offices in Cyprus. More than 30 volunteers participated in the study, which involved viewing videos over different services and networks, under regular and exceptional conditions. The participants were first introduced to the trial. Then, participants viewed HLS-based content first over IP and then over POINT, with exceptional conditions occurring during each test. Afterwards, they completed questionnaires to assess their experience. The same process was repeated with IPTV-based content. A final exit interview was conducted, before concluding the trial. During the trial, the network was monitored, gathering a wealth of information for internal analysis. However, the focus of the closed trial was assessing the user-level \emph{Quality of Experience}~(QoE), which requires direct interaction with the users.

Since the performance of PrimeTel's network and services is already of production quality, under normal operating conditions POINT simply had to match this behavior. The objective of the closed trial was rather to demonstrate that under exceptional network conditions POINT can result in a \emph{better} experience for the viewer in terms of perceived QoE. We subjected viewers into exceptional conditions with HLS and IPTV, both over the IP and the POINT network, and assessed both the objective performance of the network and the subjective evaluation of the service by the users. All sessions were recorded on video and various interesting events (e.g., video artifacts, noticeable viewer behavior, etc.) were logged.

For the HLS service, we used two widely different bitrates so that the viewer would become more aware of any drops in quality. To simulate an exceptional event we brought the HLS server down for a few seconds. There were two HLS servers in the network, main and surrogate. In the case of IP the failure led to pixelation on the screen for a long duration. Adding another HLS server on the IP network does not help, as the client must time out waiting for the primary server before switching to a different IP address for the same DNS name. Bringing the HLS server back, the player gradually returned to the higher bitrate, restoring high-quality viewing. In the ICN case, clients switched automatically to the surrogate server, since it was closer to the client than the primary server, without any noticeable effect to the end-users. When we brought back the primary server, clients continued being served by the closest server, again with no effects on viewing experience.

For the IPTV service, we exploited the fact that there are two links between the switches, but while in the IP case the spanning tree protocol uses only one link, using the other as a backup, in POINT both links are active all the time. In the IP case, we brought down the primary interface, which led to recalculation of the spanning tree and re-establishment of the IGMP snooping state, causing major viewing disruption. When the primary interface was brought back up, the same steps were repeated, leading to another service disruption. In the ICN case, the failure of the primary link led to seamless switchover to the backup link, while bringing back the primary link led to another seamless switchover to that; in both cases, there were no noticeable disruptions in the service.

\begin{figure}
\centering
\includegraphics[width=3.00in]{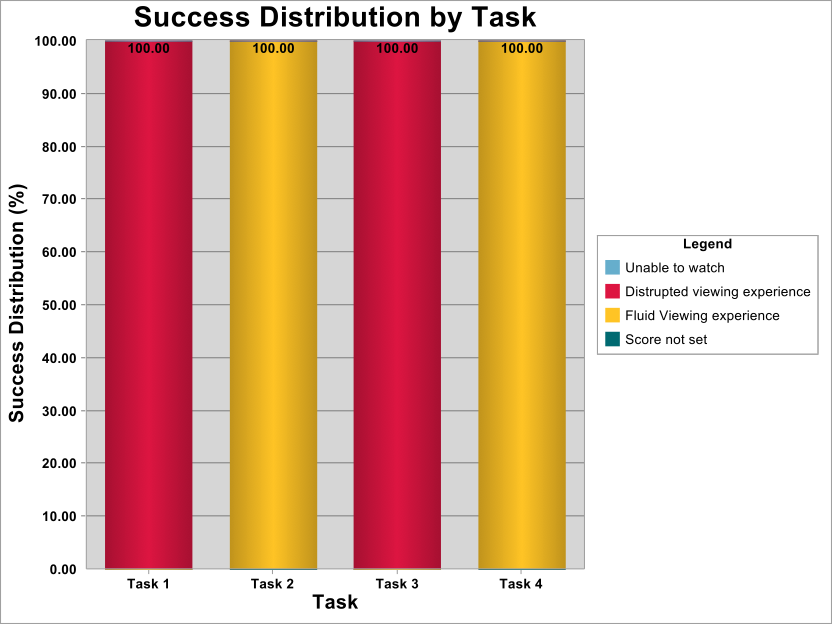}
\caption{Overall experience in each task for a random user where Tasks 1 and 3 were for an IP core, Tasks 2 and 4 were for an ICN core.}\label{success}
\end{figure}

We are currently analysing the data from the closed trials. However, in the interviews we noticed a consistent pattern of positive experiences with video over POINT. We used an adapted version of the i-QoE questionnaire~\cite{i-qoe} to assess the perceived QoE and the improved perception of content seems to be repeated in the users' answers. Figure~\ref{success} shows the responses of a random user on the viewing experience of the IPTV service when performing 4 tasks. The user found that the tasks performed over IP (1 and 3) led to a disrupted viewing experience, while with POINT (2 and 4) the viewing experience was fluid. 

\begin{figure}
\centering
\includegraphics[width=3.00in]{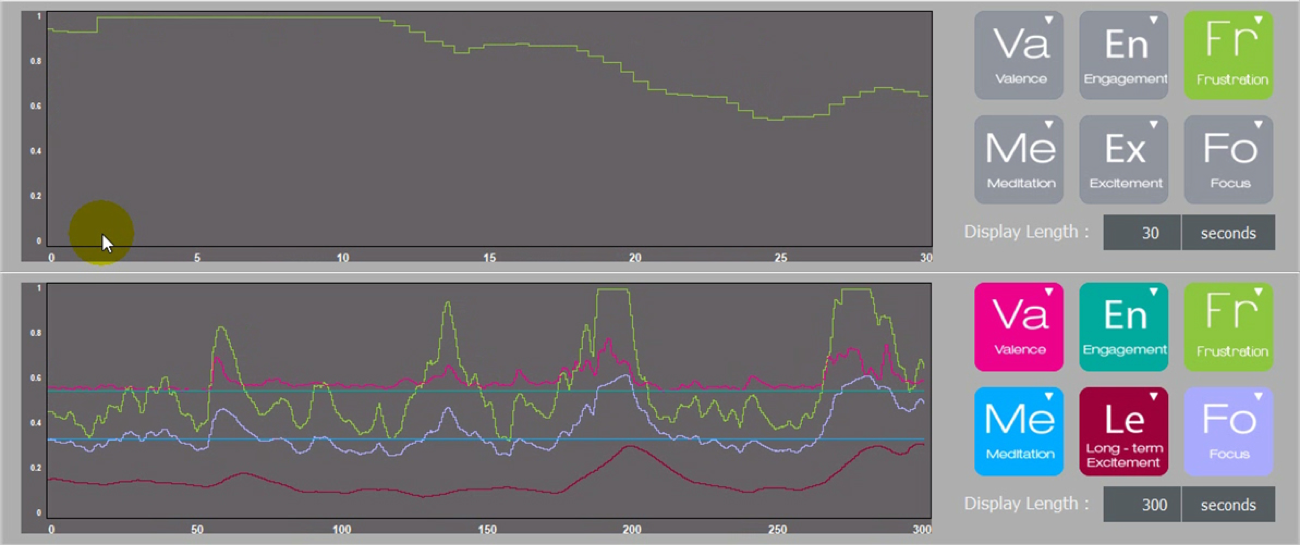}
\caption{Sample output of the EEG analyser.}\label{eeg}
\end{figure}

We also conducted a short study with the EEG analyser to record the frustration levels of users during the exceptional conditions using each networking delivery mechanism. The data followed the same pattern observed in the interviews and questionnaires. Figure~\ref{eeg} shows the output of the EEG during a test using the IP network, indicating the levels of frustration over time. Generally, users were more frustrated with content viewed on IP compared to POINT, but as it was difficult to extract exact measurements of frustration with the software we used in Cyprus, we recently repeated the study re-creating the same viewing conditions we had in Cyprus. 

\section{Conclusion and future work}

Although implementing IP services over ICN is not a novel proposition, supporting any kind of IP-based service in an operator network differentiates POINT from other ICN projects, representing a novel way to introduce ICN to the Internet. The closed trial in PrimeTel's network was the first one for an ICN project, showing that HLS and IPTV-based video services over POINT can improve upon IP under exceptional conditions common in operator networks, such as server and link congestion and failure, while maintaining equal quality during normal operation. The trial also showed that the POINT prototype is stable enough to use on a real network, with actual applications and unpredictable users.

The final step in the POINT project is to conduct an open trial, which is ongoing. The open trial will take place in actual user homes, using the same equipment and services that PrimeTel uses, and will run for two weeks, allowing us to gather large amounts of operational data on network and service performance. The questionnaires gathered from the participating users will provide additional insights on the quality of experience offered by POINT over extended periods of time. Finally, the actual process of setting up the trial has resulted in work to simplify the configuration of the POINT network, leading to numerous improvements in the management and deployment aspects of the platform.

\section*{Acknowledgments}
This research was supported by the EU funded H2020 ICT project POINT under contract 643990, and the RC-AUEB funded ``Original Scientific Publications'' project under contract ER-2766-01.

% Generated by IEEEtran.bst, version: 1.14 (2015/08/26)

\end{document}